\documentclass[prl,twocolumn,showpacs,floatfix,amsmath,amsfonts]{revtex4}
\usepackage{amsmath}
\usepackage{epsfig}
\usepackage{amssymb}
\begin{document}

\title{Feshbach resonance induced shock waves in Bose-Einstein condensates}

\author{V\'{\i}ctor M. P\'erez-Garc\'{\i}a}
\affiliation{Departamento de Matem\'aticas, Escuela T\'ecnica
Superior de Ingenieros Industriales, \\
Universidad de Castilla-La Mancha 13071 Ciudad Real, Spain}

\author{Vladimir V. Konotop}
\affiliation{Centro de F\'{\i}sica Te\'orica e Computacional,
Universidade de Lisboa, Complexo Interdisciplinar, Av. Prof. Gama
Pinto 2, Lisboa 1649-003, Portugal}

\author{Valeriy A. Brazhnyi}
\affiliation{Centro de F\'{\i}sica Te\'orica e Computacional,
Universidade de Lisboa, Complexo Interdisciplinar, Av. Prof. Gama
Pinto 2, Lisboa 1649-003, Portugal}

\begin{abstract}
We propose a method for generating shock waves in Bose-Einstein
condensates by rapidly increasing the value of the nonlinear
coefficient using Feshbach resonances. We show that in a
cigar-shaped condensate there exist primary (transverse) and secondary
(longitudinal) shock waves. We analyze how the shocks are generated in
 multidimensional scenarios and describe the phenomenology associated
 with the phenomenon.
\end{abstract}

\pacs{03.75. Lm, 03.75.Kk, 03.75.-b}
 \maketitle


Shock waves have been thoroughly investigated during the last
century in several branches of physics including fluid mechanics, plasma physics,
astrophysics, optics and solid state physics (see e.g. \cite{meyers}).
Very recently it has been recognized that there exists one more type of
media --  ultracold dilute alkali gases in a condensed state, i.e. Bose-Einstein
condensates (BEC's) -- where existence of shock waves is possible~\cite{SSS02,BKK,KZ,preprints,kamch}.

In general, BEC's represent a universal laboratory for nonlinear
phenomena. In the last few years there have been theoretical
predictions and experimental realizations of many nonlinear
structures in BEC's such as vortices, vortex lattices and vortex
rings \cite{Sols}, bright \cite{solitons} and dark \cite{dark}
solitons,  collapsing waves \cite{collapse} and stabilized
solitons \cite{Ueda}, to cite a few examples.

Theoretical studies of shock waves in BEC's so far have been
restricted to effectively one-dimensional (1D) models and explored
regimes of free expansion of the condensate.  Moreover, the
initial conditions considered in some of the previous studies seem
to be not experimentally feasible. In this Letter we propose and
investigate a new mechanism for the generation of \emph{matter
shock waves} in BEC's which could be easy to implement
experimentally. As a key ingredient of our analysis we keep the
multidimensional character of the condensate, something which is
missing in previous works \cite{SSS02,BKK,KZ,preprints,kamch}.

\emph{Estimates for shock wave generation.-} In order a shock
could develop from an initially smooth pulse, the quantum pressure
must be negligible at the initial stages of evolution. In this
situation the hydrodynamical approach (see e.g. \cite{Pit}) holds
and exploiting the analogy one can expect the appearance of a
breaking point and the subsequent development of a shock wave in a
BEC. In other words, for the breaking point to be reached one has
to require $|(\nabla^2\sqrt{n})/\sqrt{n}|\ll 8\pi a_sn$ where $n$
is the density of the condensate and $a_s$ is the s-wave
scattering length (considered positive hereafter). The quantum
pressure can be estimated as $1/L_0^2$ where $L_0$ is a
characteristic scale of the condensate to be estimated as the
minimal scale between all the characteristic spatial scales
 of the order parameter and its first
derivative. This scale must be larger than the healing length,
$\xi=(8\pi na_s)^{-1/2}$. The hydrodynamic approximation and thus
the previous estimates fail when discontinuities appear since when
a shock starts to develop $L_0 \rightarrow 0$ because of growth of
the first derivative near the edge of the shock (although the
typical size of the condensate is preserved). Due to the role of
the kinetic energy  near the shock edge one should not expect
formation of real discontinuities (like ones analyzed, for example
in \cite{KZ}) in a matter-wave system. This is similar to what it
happens in real fluids where the existence of dissipation smoothes
out the shock waves. When the characteristic size of the wave
front becomes of order of the healing length, thus making the
dispersive term relevant, we  expect the appearance of
oscillations of the wave front near the quasi-discontinuity. This
behavior, which differs from the fluid case, is due to the
essential differences between the regularization mechanisms of
fluids (diffusion) and matter waves (dispersion).

The condition $L_0^2a_sn\gg 1$ is not sufficient for the
generation of shock waves since there are two essential factors
not accounted for by this estimate. First, the density, and thus
the sound velocity rapidly decrease during the condensate
expansion. Second, the largest spatial derivative of the
condensate is not achieved where the sound velocity has its
maximum. As an example in the well known Thomas-Fermi
approximation this point is reached at the minimum of the sound
velocity. This requires special, with finite support \cite{SSS02}
or non-smooth \cite{kamch}, initial states of the condensate, in
order to produce shock waves.

Our shock-wave generation method is based on the use of Feshbach
resonances (FR) ~\cite{Feshbach0} to control the nonlinear
interactions. The idea is to make $a_s$ increase with time thus
increasing the time domain in which the quantum pressure is
negligible. In order to make qualitative estimates of the effects
expected in this scenario we assume that initially the product
$a_s n_{\text{max}}$ is negligible  so that the initial spatial
distribution of the atoms is well approximated by a gaussian
distribution. For simplicity, we consider a cylindrically
symmetric trap, with transversal and longitudinal frequencies
$\omega_\bot$ and $\omega_\|$, respectively. Then the
characteristic spatial scales
 of the problem can be identified as the healing length $\xi$, and
  the transverse $L_\bot=\sqrt{\hbar/m\omega_\bot}$ and longitudinal $L_\|=\sqrt{\hbar/m\omega_\|}$
linear oscillator lengths.
\begin{figure}
\epsfig{file=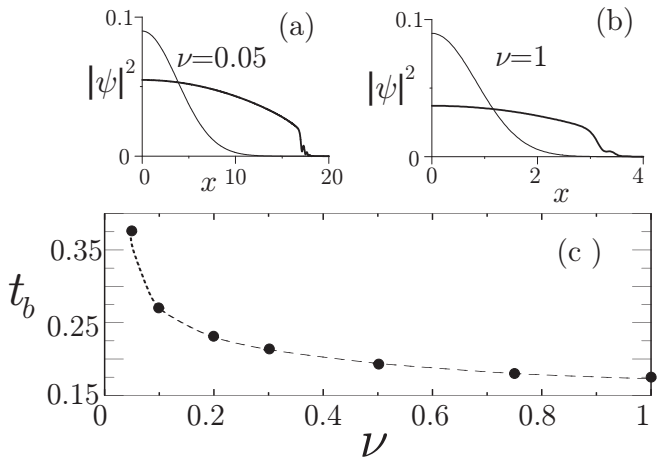,width=\columnwidth} \caption{(a,b) Initial
($t=0$, thin lines) and final ($t=t_{f}$, thick lines) profiles of a quasi-1D condensate
for $\nu=0.05$ ($t_{f}=0.45$) and $\nu=1$ ($t_{f}=0.21$). (c)
Dependence of the breaking time, $t_b$, of the 1D shock wave on the aspect
ratio $\nu$ for $T_0=0.16$ and $\tau_0=0.02$. \label{fig}}
\end{figure}

As the s-wave scattering length is ramped up to a larger value we
get an expansion of the initial state with a velocity gradient
strengthened by the parabolic potential. If the characteristic
time, $\tau_0$, of the growth of $a_s$ is small enough and the
increase of $a_s$ is large enough then at the initial stages one
can consider free expansion of the condensate. In our case there
are two characteristic velocities of the expansion \cite{nota2}
 $v_{\alpha}=\hbar/(2mL_\alpha)$ were $\alpha$ stands for $\bot$ or $\|$
 depending which direction is considered.
There are four characteristic times in this problem: the two
breaking times $t_\alpha=L_\alpha/c_{max}$ (i.e. the times at
which the breaking would occur if expansion would be absent) and
the expansion times $T_\alpha=L_\alpha/v_\alpha$. We notice that
these estimates involve $c_{max}$ which is the maximal sound
velocity after the growth of the scattering length is terminated.
In order the breaking point to be reached experimentally in a
given direction one should have  $t_\alpha\ll T_\alpha$ what can
be rewritten as the condition
$R_{\alpha}=L_{\alpha}^2a_sn_{max}\gg 1$ where $n_{max}$ is the
BEC's density in the center of the trap after the increase of the
scattering length \cite{comment}. At this point it is important to
emphasize that due to existence of the trapping potential only the
regions of large density undergo a quasi-free expansion while the
expansion of the low-density regions located near to the periphery
of the atomic cloud is suppressed by the potential. This is in a
sharp contrast to the truly  free expansion considered
elsewere~\cite{SSS02,BKK,KZ,preprints,kamch}. Our estimates
suggest that there should exist two types of shock waves in a
cigar-shaped condensate, where $t_{\bot}\ll t_\|$. A \emph{first
shock wave} would be developed in the \emph{transverse direction}
while a \emph{second shock wave} is expected to arise in the
\emph{longitudinal direction}.

\emph{The model.-} Our model is the usual time-dependent
Gross-Pitaevskii (GP) equation
\begin{equation}
\label{NLS}i\frac{\partial\psi}{\partial t} = -\frac{1}{2}
\triangle \psi + \frac{1}{2} \left( \nu^2z^2 + x^2 + y^2\right)
\psi + U|\psi|^{2}\psi,
\end{equation}
The adimensional quantities used are related with the physical
spatial ($r_j$) and temporal ($\tau$) variables through $x_j =
r_j/L_\bot$, $t = \omega_\bot\tau$. The new wavefunction $\psi$ is
related to the physical one $\Psi$ by $\psi({\bf x},t) \equiv
\Psi({\bf r},\tau)\sqrt{L_\bot^3/N}$, where $N
$ is
the number of particles in the condensate. Finally $U = 4\pi N a_s
/L_\bot$, $a_s$ being the scattering length of the cold collisions
within the condensate and $\nu = \omega_\|/\omega_{\perp}$. The
normalization for $\psi$ is $\int |\psi|^2 d^3x = 1$.
\begin{figure}
\epsfig{file=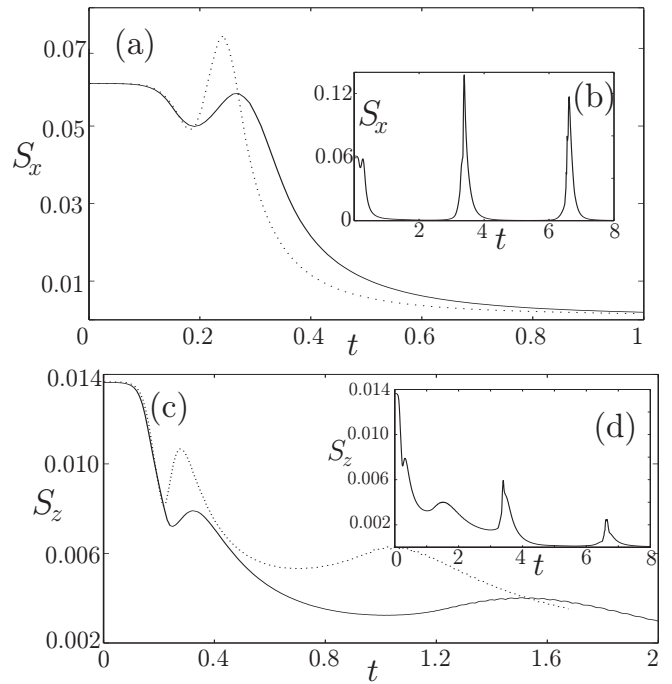,width=\columnwidth} \caption{Results of 2D
simulations for $\tau_0 = 0.02, \nu = 0.05$, and grid parameters
$\Delta t = 0.001, N_x = 2048, N_z = 1024$ on $(x,y) \in
[-30,30]\times[-60,60]$. (a) Short and (b) long time evolution of
the derivatives $S_x(t) = \max_{x,z}|\partial \psi/\partial x|$.
(c) Short and (d) long time evolution of the derivatives
$S_z(t) = \max_{x,z}|\partial \psi/\partial z|$. Dotted lines correspond
to $T_0 = 0.16$ while solid lines are for $T_0 = 0.14$.
\label{prima}}
\end{figure}

We ramp up the scattering length according to
\begin{equation}
U(t) = \begin{cases} \alpha\left(e^{t/\tau_0}-1\right), & t<T_0, \\
\alpha\left(e^{T_0/\tau_0}-1\right) & t \geq T_0.
\end{cases}
\label{Fesch}
\end{equation}
In what follows we choose $\alpha = 1$ without loss of generality
and study the dependence of the formation of shock waves on the
parameters $T_0, \tau_0, \nu$.

\emph{One-dimensional shocks.-} We first consider the
cross-section of the condensate along one of
 the directions (radial or axial) as an effectively 1D case. By solving
 numerically the 1D version of the GPE (\ref{NLS}) we have
 found the dependence of the time of the shock wave formation $t_b$ on the
 size of the condensate characterized by the trap asymmetry. As it clear
 from Fig.\ref{fig} the shock waves develop faster in
time in systems with smaller aspect ratio in agreement with our
previous qualitative arguments.

\begin{figure}
\epsfig{file=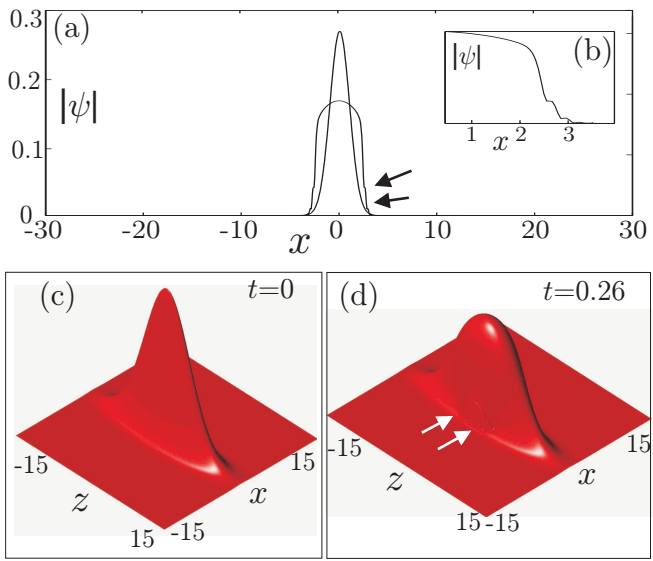,width=\columnwidth} \caption{Formation of
the first shock along the $x$ direction for parameters $T_0 =
0.16, \tau_0 = 0.02, \nu = 0.05$. (a) $|\psi(x,z)|$ for $t=0$ and
$t=0.26$ just after the first shock has formed. (b) Detail of the
region $x\in[1,3]$ where the irregularities in which the shock
degenerates are more evident. (c-d) Surface plots of $|\psi(x,z)|$
for (c) $t=0$ and (d) $t=0.26$. In all cases the arrows mark the
location of the shock wave.\label{segunda}}
\end{figure}

\begin{figure}
\epsfig{file=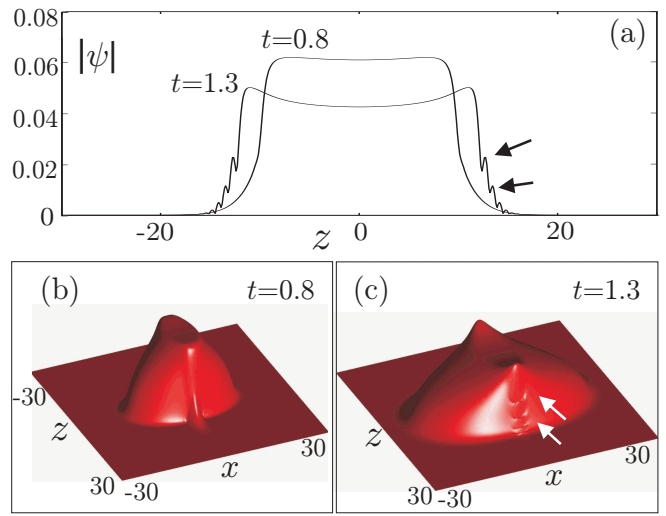,width=\columnwidth}\caption{Formation of
the first shock along the $z$ direction for parameters $T_0 =
0.16, \tau_0 = 0.02, \nu = 0.05$. (a) $|\psi(x,z)|$ for $t=0.8$
and $t=1.3$ after the shock has formed.(c-d) Surface plots of
$|\psi(x,z)|$ for (c) $t=0.8$ and (d) $t=1.3$. In all cases the
arrows mark the location of the shock wave.\label{tercera}}
\end{figure}

\emph{Two-dimensional shocks.-} We have studied the formation of
shock waves from gaussian ground states under variation of the
parameters $\nu,\tau_0$, and $T_0$. Results of typical numerical
simulations of Eq. \eqref{NLS} are summarized in Figs.
\ref{prima},\ref{segunda} and \ref{tercera} corresponding to $T_0
= 0.14, \tau_0 = 0.02$ thus $ U_{\text{max}} = U(T_0) \simeq 1100$
[Fig. \ref{prima} solid lines] and $T_0 = 0.16, \tau_0 = 0.02$,
$U_{\text{max}} \simeq 3 \times 10^3$ [Fig. ~\ref{prima}(a,c)
dotted lines and Figs. ~\ref{segunda} and \ref{tercera}]. The
increase of the nonlinearity leads to an expansion of the
condensate which is faster along the transverse direction due to
the higher compression and thus the increased non-stationarity
along this direction (notice that this is the case $t_\bot\ll
t_\|$ qualitatively described above, see also \cite{BKK}). On long
time scales the effect of the potential is to confine the
wavepacket thus generating recurrent oscillations.

\begin{figure}
\epsfig{file=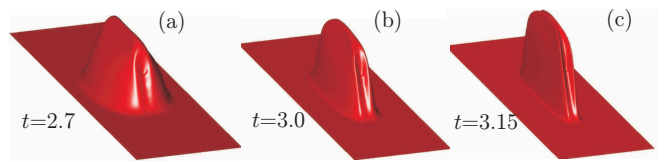,width=\columnwidth}\caption{Surface plots
of $|\psi(x,z)|$ for (a) $t=2.7$, (b) $t=3.0$ and (c) $t = 3.5$
showing the formation of strong shocks for long times due to the
oscillations of the wavepacket width in the harmonic potential.
\label{cuarta}}
\end{figure}

As it has been predicted above qualitatively and on the basis of
the numerical simulations of the 1D model, the faster transversal
expansion implies that the shock wave is first generated along the
\emph{transverse} direction and not along the longitudinal one as
one could naively think because of the geometry of the condensate.
The formation of this shock is seen as an increase of the maximum
value of the spatial derivatives along $x$ [Fig. \ref{prima}(a)].
This phenomenon is accompanied by a small sympathetic increase on
the derivatives along $z$ which does not correspond to a true
shock. The true longitudinal shock appears later due to the slower
expansion along this direction [Fig. \ref{prima}(c)].

The structures of the longitudinal and transverse shocks are shown
in Figs. \ref{segunda} and \ref{tercera}.
 It can be seen how the longitudinal shocks lead to a very
 oscillating behavior near the shock edge while for the transverse shocks
  the behavior is smoother, although small amplitude
 oscillations are also present.

 For longer times and due to the effect of the trapping potential
 we observe the recurrent formation of very strong shocks (see the peaks in $S_x,S_z$ for $t\simeq \pi, 2\pi$ in
 Fig. \ref{prima}). In Fig. \ref{cuarta} we observe that the wave becomes filamented due to the
  strongly compression induced by the combined effect
of nonlinear forces and the harmonic potential, a phenomenon which
could be used as a very efficient matter wave compressor.

\begin{figure}
\epsfig{file=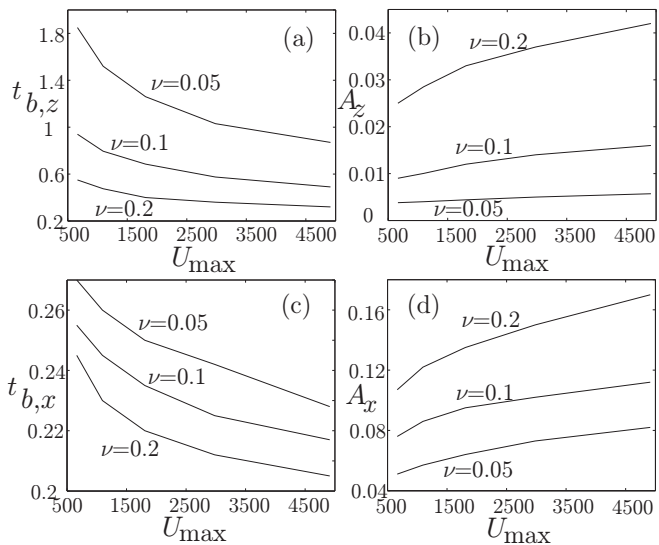,width=\columnwidth} \caption{Amplitude
 and breaking time of the
longitudinal and transverse shocks as a function of
$U_{\text{max}}$ for different spatial asymmetries. $A_{x} =
S_{x}(t_{b,x})$ and $A_{z} = S_{z}(t_{b,z})$ with $S_{x}$ and
$S_z$ as in Fig. 2. \label{quinta}}
\end{figure}

The amplitude of the shocks increases as the nonlinearity is
increased. In Fig. \ref{quinta} we study the dependence of the
shocks amplitude and breaking times as a function of the trap
asymmetry $\nu$. It can be seen that as the trap becomes more
symmetric the breaking time for the transverse and longitudinal
shocks become more similar. This is a reasonable result since in
the case $\nu = 1$ the shocks should develop at the same time
along all the spatial directions starting from symmetric initial
data. We have also studied the dependence of the shock formation
on the rising time of the nonlinearity $\tau_0$ and seen that it
plays an essential role in the transverse shock formation. For
instance, taking $\tau_0 = 0.1$ and $T_0 = 0.70$, which give the
same  values of $U_{\text{max}}$ as those shown in Figs.
\ref{prima},\ref{segunda} and \ref{tercera}, we have found that
there are no transverse shocks (only those corresponding to the
revival of the oscillation in the trap appear). This effect has a
simple explanation: since $L_\bot\ll L_\|$, for a definite region
of densities
 $n_{max}$ one may have $R_\|\gg R_\bot\sim 1$, what means that it
 is possible to reach the breaking point only longitudinally.


\emph{Three-dimensional condensates.-} We have also studied the
formation of shock waves in three-dimensional condensates using
the same type of numerical methods on grids of up to
256$\times$256$\times$200 points. With this resolution  we have
been  able at least to detect the formation of the first shock
waves. We have found features similar to those described for the
two-dimensional case with the shocks first generating transversely
and later longitudinally if the nonlinearity rises fast enough.

To conclude, we have proposed a mechanism for generating shock
waves and described their dynamics. In doing so we have kept the
multidimensional nature of the system, something which is missing
in previous works and could lead to flawed results.  We hope that
our method could guide experimentalists in the field of matter
waves to observe the formation of shock waves with Bose-Einstein
condensates.

V. M. P-G. is partially supported by Ministerio de Ciencia y
Tecnolog\'{\i}a (MCyT) under grant BFM2003-02832 and
Consejer\'{\i}a de Ciencia y Tecnolog\'{\i}a de la Junta de
Comunidades de Castilla-La Mancha under grant PAC-02-002. V.V.K.
acknowledges support from the European grant, COSYC n.o.
HPRN-CT-2000-00158. Work of V.A.B. has been supported by the FCT
fellowship SFRH/BPD/5632/2001. Cooperative work has been supported
by the Integrated-Action No E-23/03 and Acciones Integradas of
MCyT HP2002-0059.


\begin{thebibliography}{9}
\bibitem{comment}{The 1D analog of this criterium was introduced in Ref.~\cite{KZ} where $R$ is a ``Reinolds''
number. In our case we have different transverse and longitudinal
Reinolds numbers.}
\end{thebibliography}
\end{document}